\documentclass[letterpaper,11pt]{article}

\usepackage[latin1]{inputenc}
\usepackage[T1]{fontenc}
\usepackage{graphicx}
\usepackage{amsfonts}
\usepackage{amssymb}
\usepackage{amsmath} 
\usepackage{amsthm}
\usepackage{url}

\usepackage{natbib}

\usepackage[margin=1in]{geometry}

\newcommand{\RR}{\mathbb{R}}

\begin{document}

\title{Kernel methods for \emph{in silico} chemogenomics}
\author{
Laurent Jacob\\
Centre for Computational Biology\\
Ecole des Mines de Paris\\
35, rue Saint-Honor\'e\\
77300 Fontainebleau, France\\
\texttt{laurent.jacob@ensmp.fr}
\and
Jean-Philippe Vert\\
Centre for Computational Biology\\
Ecole des Mines de Paris\\
35, rue Saint-Honor\'e\\
77300 Fontainebleau, France\\
\texttt{jean-philippe.vert@ensmp.fr}
}

\maketitle

\begin{abstract}
  Predicting interactions between small molecules and proteins is a
  crucial ingredient of the drug discovery process. In particular,
  accurate predictive models are increasingly used to preselect
  potential lead compounds from large molecule databases, or to screen
  for side-effects.  While classical \emph{in silico} approaches focus
  on predicting interactions with a given specific target, new
  chemogenomics approaches adopt cross-target views.  Building on
  recent developments in the use of kernel methods in bio- and
  chemoinformatics, we present a systematic framework to screen the
  chemical space of small molecules for interaction with the
  biological space of proteins. We show that this framework allows
  information sharing across the targets, resulting in a dramatic
  improvement of ligand prediction accuracy for three important
  classes of drug targets: enzymes, GPCR and ion channels.
\end{abstract}

\section{Introduction}

Predicting interactions between small molecules and proteins is a
key element in the drug discovery process. In particular,
several classes of proteins such as G-protein-coupled receptors
(GPCR), enzymes and ion channels represent a large fraction of current
drug targets and important targets for new drug
development~\citep{Hopkins2002druggable}. Understanding and predicting
the interactions between small molecules and such proteins could
therefore help in the discovery of new lead compounds.

Various approaches
have already been developed and have proved very useful to address
this \emph{in silico} prediction issue~\citep{Manly2001impact}. The
classical paradigm is to predict the modulators of a given target,
considering each target as a different problem. Usual methods are
classified into \emph{ligand-based} and \emph{structure-based} or
\emph{docking} approaches. Ligand-based approaches compare a candidate
ligand to the known ligands of the target to make their prediction,
typically using machine learning
algorithms~\citep{Butina2002Predicting,Byvatov2003Comparison,Zernov2003Drug}
whereas structure-based approaches use the 3D-structure of the target
to determine how well each candidate binds the
target~\citep{Halperin2002Principles}.

Ligand-based approaches necessitate to know enough ligands of a given
target with respect to the complexity of the ligand/non-ligand
separation to produce accurate predictors. If few or no ligands are
known for a target, one is compelled to use docking approaches, which
in turn necessitate to know the 3D structure of the target and are
very time consuming. If for a given target with unavailable 3D
structure no ligand is known, none of the classical approaches can
apply. This is the case for many GPCR as very few structures have been
crystallized so far~\citep{Ballesteros2001G} and many of these
receptors, referred to as \emph{orphan} GPCR, have no known ligand.

An interesting way to solve this problem is to cast it in the
\emph{chemogenomics} framework. Chemogenomics aims at mining the
\emph{chemical space}, which roughly corresponds to the set of all
small molecules, for interactions with the \emph{biological space},
i.e., the set of all proteins, in particular drug targets. A salient
feature of the chemogenomics approach is the realization that some
classes of molecules can bind ``similar'' proteins, suggesting that
the knowledge of some ligands for a target can be helpful to determine
ligands for similar targets. Besides, this type of method allows for a
more rational approach to design drugs since controlling a whole
ligand's selectivity profile is crucial to make sure that no side
effect occurs and that the compound is compatible with therapeutical
usage.

Recent
reviews~\citep{Kubinyi2004Chemogenomics,2006Chemical,Klabunde2007Chemogenomic,Rognan2007Chemogenomic}
list several chemogenomic approaches to predict interactions between
compounds and targets~\citep{Oloff2006Chemometric,Bock2005Virtual}.
Many of these chemogenomics methods rely on some fixed choice of which
targets should be used when learning a predictor for a given target,
the most extreme example being the learning of a predictor for a whole
family or subfamily of
targets~\citep{Balakin2002Property-based,Klabunde2006ChemogenomicsA}.
Most of them also need some specific procedure to choose which ligands
of the selected targets are used and how they are used. 

We propose a method that uses existing and well tested machine
learning algorithms, casting the interaction prediction problem in a
joint ligand-target space. This embeds the sharing level threshold
problem in a simple representation choice for which we also propose a
systematic approach based on combinations of features of the ligand
and features of the target. For the three families of targets of
interest, we show that our approach outperforms the state-of-the-art
individual SVM, and gives good performances even for targets with no
known ligand.

\section{Method}

We formulate the typical \emph{in silico} chemogenomics problem as the
following learning problem: given a collection of $n$ target/molecule
pairs $(t_{1},c_{1}), \ldots(t_{n},c_{n})$ known to interact or not,
estimate a function $f(t,c)$ that would predict whether any chemical
$c$ binds to any target $t$. In this section we propose a rigorous and
general framework to solve this problems, building on recent
developments of kernel methods in bio- and chemoinformatics.

\subsection{From single-target screening to chemogenomics}

Much effort in chemoinformatics has been devoted to the more
restricted problem of mining the chemical space for interaction with a
single target $t$, using a training set of molecules
$c_{1},\ldots,c_{n}$ known to interact or not with the target. Machine
learning approaches, such as artificial neural networks (ANN) or
support vector machines (SVM), often provide competitive models for
such problems.  The simplest linear models start by representing each
molecule $c$ by a vector representation $\Phi(c)$, before estimating a
linear function $f_{t}(c) = w_{t}^\top \Phi(c)$ whose sign (positive
or negative) is used to predict whether or not the small molecule $c$
is a ligand of the target $t$. The weight vector $w_{t}$ is typically
estimated based on its ability to correctly predict the classes of
molecules in the training set.

The \emph{in silico} chemogenomics problem is more general because
data involving interactions with different targets are available to
train a model which must be able to predict interactions between any
molecule and any protein. In order to extend the previous machine
learning approaches to this setting, we need to represent a
\emph{pair} $(t,c)$ of target $t$ and chemicals $c$ by a vector
$\Phi(t,c)$, then estimate a linear function $f(t,c) = w^\top
\Phi(t,c)$ whose sign is used to predict whether or not $c$ can bind
to $t$. As before the vector $w$ can be estimated from the training
set of interacting and non-interacting pairs, using any linear machine
learning algorithm.

To summarize, we propose to cast the \emph{in silico} chemogenomics
problem as a learning problem in the ligand-target space thus making
it suitable to any classical linear machine learning approach as soon
as a vector representation $\Phi(t,c)$ is chosen for protein/ligand
pairs.  We propose in the next sections a systematic way to design
such a representation.

\subsection{Vector representation of target/ligand pairs}

A large literature in chemoinformatics has been devoted to the problem
of representing a molecule $t$ by a vector $\Phi_{ligand}(c) \in
\RR^{d_{c}}$, e.g., using various molecular
descriptors~\citep{Todeschini2002Handbook}. These descriptors encode
several features related to the physico-chemical and structural
properties of the molecules, and are widely used to model interactions
between the small molecules and a single target using linear models
described in the previous
section~\citep{Gasteiger2003Chemoinformatics}.  Similarly, much work
in computational biology has been devoted to the construction of
descriptors for genes and proteins, in order to represent a given
protein $t$ by a vector $\Phi_{target}(t) \in \RR^{d_{t}}$. The
descriptors typically capture properties of the sequence or structure
of the protein, and can be used to infer models to predict, e.g., the
structural or functional class of a protein.

For our \emph{in silico} chemogenomics problem we need to represent
each pair $(c,t)$ of small molecule and protein by a single vector
$\Phi(c,t)$. In order to capture interactions between features of the
molecule and of the protein that may be useful predictors for the
interaction between $c$ and $t$, we propose to consider features for
the pair $(c,t)$ obtained by multiplying a descriptor of $c$ with a
descriptor of $t$. Intuitively, if for example the descriptors are
binary indicators of specific structural features in each small molecule and proteins,
then the product of two such features indicates that both the small
molecule and the target carry specific features, which may be strongly
correlated with the fact that they interact. More generally, if a molecule $c$ is
represented by a vector of descriptors $\Phi_{ligand}(c) \in
\RR^{d_{c}}$ and a target protein by a vector of descriptors
$\Phi_{target}(t) \in \RR^{d_{t}}$, this suggests to represent the pair
$(c,t)$ by the set of all possible products of features of $c$ and
$t$, i.e., by the tensor product:
\begin{equation}\label{eq:tp}
\Phi(c,t) = \Phi_{ligand}(c) \otimes \Phi_{target}(t)\,.
\end{equation}
Remember that the tensor product in \eqref{eq:tp} is a $d_{c}\times
d_{t}$ vector whose $(i,j)$-th entry is exactly the product of the
$i$-th entry of $\Phi_{ligand}(c)$ by the $j$-th entry of
$\Phi_{target}(t)$. This representation can be used to combine in a
principled way any vector representation of small molecules with any
vector representation of proteins, for the purpose of \emph{in silico}
chemogenomics or any other task involving pairs of molecules/protein.
A potential issue with this approach, however, is that the size of the
vector representation for a pair may be prohibitively large for
practical computation and storage. For example, using a vector of
molecular descriptors of size $1024$ for molecules and representing a
protein by the vector of counts of all $2$-mers of amino-acids in its
sequence ($d_{t}=20\times 20 = 400$) results in more than 400k
dimensions for the representation of a pair. In order to circumvent
this issue we now show how kernel methods such as SVM can efficiently
work in such large spaces.

\subsection{Kernels for target/ligand pairs}

SVM is an algorithm to estimate linear binary classifiers from a
training set of patterns with known
class~\citep{Boser1992training,Vapnik1998Statistical}. A salient
feature of SVM, often referred to as the \emph{kernel trick}, is its
ability to process large- or even infinite-dimensional patterns as
soon as the inner product between any two patterns can be efficiently
computed.  This property is shared by a large number of popular linear
algorithms, collectively referred to as \emph{kernel methods},
including for example algorithms for regression, clustering or outlier
detection~\citep{Scholkopf2002Learning,Shawe-Taylor2004Kernel}.

In order to apply kernel methods such as SVM for \emph{in silico}
chemogenomics, we therefore need to show how to efficiently compute
the inner product between the vector representations of two
molecule/protein pairs. Interestingly, a classical and easy to check
property of tensor products allows to write the inner product between
two tensor product vectors as a product of inner products:
\begin{equation}\label{eq:inptp}
  \left(\Phi_{ligand}(c) \otimes \Phi_{target}(t)\right)^\top \left(\Phi_{ligand}(c') \otimes \Phi_{target}(t')\right) = \Phi_{ligand}(c)^\top \Phi_{ligand}(c') \times  \Phi_{target}(t)^\top \Phi_{target}(t')\,.
\end{equation}
This factorization dramatically reduces the burden of working with
tensor products in large dimensions. For example, in our previous
example where the dimensions of the small molecule and proteins are
vectors of respective dimensions $1024$ and $400$, the inner product
in $>400k$ dimensions between tensor products is simply obtained from
(\ref{eq:inptp}) by computing two inner products, respectively in
dimensions $1024$ and $400$, before taking their product.

Even more interestingly, this reasoning extends to the case where
inner products between vector representations of small molecules and
proteins can themselves be efficiently computed with the help of
\emph{positive definite} kernels~\citep{Vapnik1998Statistical}, as
explained in the next sections.  Positive definite kernels are linked
to inner products by a fundamental result~\citep{Aronszajn1950Theory}:
the kernel between two points is equivalent to an inner product
between the points mapped to a Hilbert space uniquely defined by the
kernel. Now by denoting
\begin{displaymath}
  K_{ligand}(c,c') = \Phi_{ligand}(c)^\top \Phi_{ligand}(c')\,,\quad  K_{target}(t,t') =\Phi_{target}(t)^\top \Phi_{target}(t')\,,
\end{displaymath}
we obtain the inner product between tensor products by:
\begin{equation}
\label{eq:product}
K\left((c,t),(c',t')\right) = K_{target}(t,t')\times K_{ligand}(c,c').
\end{equation}

In summary, as soon as two kernels $K_{ligand}$ and $K_{target}$
corresponding to two implicit embeddings of the chemical and
biological spaces in two Hilbert spaces are chosen, we can solve the
\emph{in silico} chemogenomics problem with an SVM (or any other
relevant kernel method) using the product kernel (\ref{eq:product})
between pairs. The particular kernels $K_{ligand}$ and $K_{target}$
should ideally encode properties related to the ability of similar
molecules to bind similar targets or ligands respectively. We review
in the next two sections possible choices for such kernels.

\subsection{Kernels for ligands}
Recent years have witnessed impressive advances in the use of SVM in
chemoinformatics~\citep{Ivanciuc2007Applications}. In particular much
work has focused on the development of kernels for small molecules for
the purpose of single-target virtual screening and prediction of
pharmacokinetics and toxicity. For example simple inner products
between vectors of classical molecular descriptors have been widely
investigated, including physicochemical properties of molecules or 2D
and 3D fingerprints~\citep{Todeschini2002Handbook,Azencott2007One}.
Other kernels have been designed directly from the comparison of 2D
and 3D structures of molecules, including kernels based on the
detection of common substructures in the 2D structures molecules seen
as
graphs~\citep{Kashima2003Marginalized,Kashima2004Kernels,Gartner2003graph,
  Mahe2005Graph,Ralaivola2005Graph,Borgwardt2005Shortest-Path,
  Ramon2003Expressivity,Horvath2004Cyclic,Mahe2006Graph} or on the
encoding of various properties of the 3D structure of a
molecules~\citep{Mahe2006Pharmacophore,Azencott2007One}.

While any of these kernels could be used to model the similarities of
small molecules and be plugged into (\ref{eq:product}), we restrict
ourselves in our experiment to a particular kernel proposed
by~\cite{Ralaivola2005Graph} called the \emph{Tanimoto kernel}, a
classical choice that usually gives state-of-the-art performances in
molecule classification tasks. It is defined as:
\begin{equation}
\label{eq:tanimoto}
K_{ligand}(c,c') =
\frac{\Phi_{ligand}(c)^\top\Phi_{ligand}(c')}{\Phi_{ligand}(c)^\top\Phi_{ligand}(c)+\Phi_{ligand}(c')^\top\Phi_{ligand}(c')-\Phi_{ligand}(c)^\top\Phi_{ligand}(c')}\,,
\end{equation}
where $\Phi_{ligand}(c)$ is a binary vector whose bits indicate the
presence or absence of all linear path of length $l$ or less as
subgraph of the 2D structure of $c$. We chose $l=8$ in our experiment,
i.e., characterize the molecules by the occurrences of linear
subgraphs of length $8$ or less, a value previously observed to give
good results in several virtual screening task \cite{Mahe2005Graph}.
We used the freely and publicly available
\emph{ChemCPP}\footnote{Available at
  \url{http://chemcpp.sourceforge.net}.} software to compute this
kernel in the experiments.

\subsection{Kernels for targets}

SVM and kernel methods are also widely used in
bioinformatics~\citep{Schoelkopf2004Kernel}, and a variety of
approaches have been proposed to design kernels between proteins,
ranging from kernels based on the amino-acid sequence of a
protein~\citep{Jaakkola2000Discriminative,Leslie2002spectrum,
  Tsuda2002Marginalized,Ben-Hur2003Remote,Leslie2004Mismatch,
  Vert2004Local,Kuang2005Profile-based,Cuturi2005context-tree} to
kernels based on the 3D structures of
proteins~\citep{Dobson2005Predicting,Borgwardt2005Protein,Qiu2007structural}
or the pattern of occurrences of proteins in multiple sequenced
genomes~\citep{Vert2002tree}. These kernels have been used in
conjunction with SVM or other kernel methods for various tasks related
to structural or functional classification of proteins. While any of
these kernels can theoretically be used as a target kernel in
\eqref{eq:product}, we investigate in this paper a restricted list of
specific kernels described below, aimed at illustrating the
flexibility of our framework and test various hypothesis.
\begin{itemize}
\item The \emph{Dirac} kernel between two targets $t,t'$ is:
\begin{equation}
  K_{Dirac}(t,t') = \begin{cases}
    1 &\text{ if }t=t'\,,\\
    0&\text{ otherwise.}
\end{cases}
\end{equation}
This basic kernel simply represents different targets as orthonormal
vectors. From (\ref{eq:product}) we see that orthogonality between two
proteins $t$ and $t'$ implies orthogonality between all pairs $(c,t)$
and $(c',t')$ for any two small molecules $c$ and $c'$. This means
that a linear classifier for pairs $(c,t)$ with this kernel decomposes
as a set of independent linear classifiers for interactions between
molecules and each target protein, which are trained without sharing
any information of known ligands between different targets. In other
words, using Dirac kernel for proteins amounts to performing classical
learning independently for each target, which is our baseline
approach.
\item The \emph{multitask} kernel between two targets $t,t'$ is
  defined as:
  \begin{displaymath}
        K_{multitask}(t,t') = 1+K_{Dirac}(t,t')\,.
  \end{displaymath}
  This kernel, originally proposed in the context of multitask
  learning \cite{Evgeniou2005Learning}, removes the orthogonality of
  different proteins to allow sharing of information. As explained in
  \cite{Evgeniou2005Learning}, plugging $K_{multitask}$ in
  \eqref{eq:product} amounts to decomposing the linear function used
  to predict interactions as a sum of a linear function common to all
  targets and of a linear function specific to each target:
  \begin{displaymath}
  f(c,t) = w^\top \Phi(c,t) = w_{general}^\top \Phi_{ligand}(c) + w_{t}^\top\Phi_{ligand}(c)\,.
  \end{displaymath}
  A consequence is that only data related to the the target $t$ are
  used to estimate the specific vector $w_{t}$, while all data are
  used to estimate the common vector $w_{general}$. In our framework
  this classifier is therefore the combination of a target-specific
  part accounting for target-specific properties of the ligands and a
  global part accounting for general properties of the ligands across
  the targets.  The latter term allows to share information during the
  learning process, while the former ensures that specificities of the
  ligands for each target are not lost.
\item While the multitask kernel provides a basic framework to share
  information across proteins, it does not allow to weight differently
  how known interactions with a protein $t$ should contribute to
  predict interactions with a target $t'$. Empirical observations
  underlying chemogenomics, on the other hand, suggest that molecules
  binding a ligand $t$ are only likely to bind ligand $t'$ similar to
  $t$ in terms of structure or evolutionary history. In terms of
  kernels this suggest to plug into~\eqref{eq:product} a kernel for
  proteins that quantifies this notion of similarity between proteins,
  which can for example be detected by comparing the sequences of
  proteins. In order to test this approach, we therefore tested two
  commonly-used kernels between protein sequences: the mismatch
  kernel~\citep{Leslie2004Mismatch}, which compares proteins in terms
  of common short sequences of amino acids up to some mismatches, and
  the local alignment kernel~\citep{Vert2004Local} which measures the
  similarity between proteins as an alignment score between their
  primary sequences. In our experiments involving the mismatch kernel,
  we use the classical choice of $3$-mers with a maximum of $1$
  mismatch, and for the datasets where some sequences were not
  available in the database, we added $K_{Dirac}(t,t')$ to the kernel
  (and normalized at $1$ on the diagonal) in order to keep it valid.
\item Alternatively we propose a new kernel aimed at encoding the
  similarity of proteins with respect to the ligands they bind.
  Indeed, for most major classes of drug targets such as the ones
  investigated in this study (GPCR, enzymes and ion channels),
  proteins have been organized into hierarchies that typically
  describe the precise functions of the proteins within each family.
  Enzymes are labeled with \emph{Enzyme Commission numbers} (EC
  numbers) defined in~\cite{International1992Enzyme}, that classify
  the chemical reaction they catalyze, forming a $4$-level hierarchy
  encoded into 4 numbers. For example $EC\;1$ includes
  oxydoreductases, $EC\;1.2$ includes oxidoreductases that act on the
  aldehyde or oxo group of donors, $EC\;1.2.2$ is a subclass of
  $EC\;1.2$ with $NAD+$ or $NADP+$ as acceptor and $EC\;1.2.2.1$ is a
  subgroup of enzymes catalyzing the oxidation of formate to
  bicarbonate. These number define a natural and very informative
  hierarchy on enzymes: one can expect that enzymes that are closer in
  the hierarchy will tend to have more similar ligands.  Similarly,
  GPCRs are grouped into $4$ classes based on sequence homology and
  functional similarity: the \emph{rhodopsin} family (class A), the
  \emph{secretin} family (class B), the \emph{metabotropic} family
  (class C) and a last class regrouping more diverse receptors (class
  D). The KEGG database~\citep{Kanehisa2002KEGG} subdivides the large
  rhodopsin family in three subgroups (amine receptors, peptide
  receptors and other receptors) and adds a second level of
  classification based on the type of ligands or known subdivisions.
  For example, the rhodopsin family with amine receptors is subdivided
  into cholinergic receptors, adrenergic receptors, \emph{etc}. This
  also defines a natural hierarchy that we could use to compare GPCRs.
  Finally, KEGG also provides a classification of ion channels.
  Classification of ion channels is a less simple task since some of
  them can be classified according to different criterions like
  voltage dependence or ligand-gating. The classification proposed by
  KEGG includes \emph{Cys-loop superfamily, glutamate-gated cation
    channels, epithelial and related Na+ channels, voltage-gated
    cation channels, related to voltage-gated cation channels, related
    to inward rectifier K+ channels, chloride channels} and
  \emph{related to ATPase-linked transporters} and each of these
  classes is further subdivided according for example to the type of
  ligands (\emph{e.g.}, glutamate receptor) or to the type of ion
  passing through the channel (\emph{e.g.}, Na+ channel). Here again,
  this hierarchy can be used to define a meaningful similarity in
  terms of interaction behavior.

  For each of the three target families, we define the hierarchy
  kernel between two targets of the family as the number of common
  ancestors in the corresponding hierarchy plus one, that is,
  \begin{displaymath}
    K_{hierarchy}(t,t') = \langle\Phi_h(t),\Phi_h(t')\rangle,
  \end{displaymath}
  where $\Phi_h(t)$ contains as many features as there are nodes in
  the hierarchy, each being set to $1$ if the corresponding node is
  part of $t$'s hierarchy and $0$ otherwise, plus one feature
  constantly set to one that accounts for the "plus one" term of the
  kernel.
  \end{itemize}

\section{Data}
We extracted compound interaction data from the KEGG BRITE
Database~\citep{Kanehisa2002KEGG,Kanehisa2004KEGG} concerning enzyme,
GPCR and ion channel, three target classes particularly relevant for
novel drug development.  

For each family, the database provides a list of known compounds for
each target. Depending on the target families, various categories of
compounds are defined to indicate the type of interaction between each
target and each compound. These are for example \emph{inhibitor,
  cofactor} and \emph{effector} for enzyme ligands, \emph{antagonist}
or \emph{(full/partial) agonist} for GPCR and \emph{pore blocker,
  (positive/negative) allosteric modulator, agonist} or
\emph{antagonist} for ion channels. The list is not exhaustive for the
latter since numerous categories exist. Although different types of
interactions on a given target might correspond to different binding
sites, it is theoretically possible for a non-linear classifier like
SVM with non-linear kernels to learn classes consisting of several
disconnected sets. Therefore, for the sake of clarity of our analysis,
we do not differentiate between the categories of compounds.

We eliminated all compounds for which no molecular descriptor was
available (principally peptide compounds), and all the targets for
which no compound was known. For each target, we generated as many
negative ligand-target pairs as we had known ligands forming positive
pairs by combining the target with a ligand randomly chosen among the
other target's ligands (excluding those that were known to interact
with the given target). This protocol generates false negative data
since some ligands could actually interact with the target although
they have not been experimentally tested, and our method could benefit
from experimentally confirmed negative couples.

\begin{figure}[ht]
  \begin{minipage}[b]{0.33\linewidth}
  \centering
  \includegraphics[width=\linewidth]{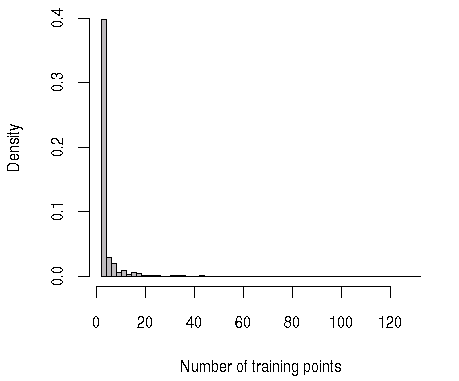}
  \end{minipage}
  \begin{minipage}[b]{0.33\linewidth}
  \centering
  \includegraphics[width=\linewidth]{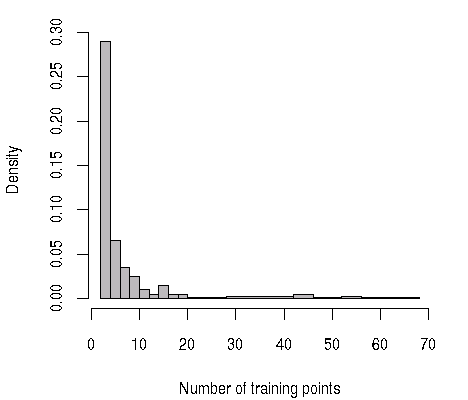}
  \end{minipage}
  \begin{minipage}[b]{0.33\linewidth}
  \centering
  \includegraphics[width=\linewidth]{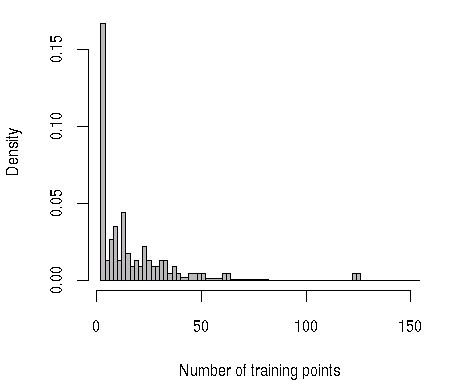}
  \end{minipage}
  \caption{Distribution of the number of training points for a target
    for the enzymes, GPCR and ion channel datasets.}
  \label{fig:distrib}
\end{figure}

This resulted in $2436$ data points for enzymes ($1218$ known
enzyme-ligand pairs and $1218$ generated negative points) representing
interactions between $675$ enzymes and $524$ compounds, $798$ training
data points for GPCRs representing interactions between $100$
receptors and $219$ compounds and $2330$ ion channel data points
representing interactions between $114$ channels and $462$ compounds.
Besides, Figure~\ref{fig:distrib} shows the distribution of the number
of known ligands per target for each dataset and illustrates the fact
that for most of them, few compounds are known.

For each target $t$ in each family, we carried out two experiments.
First, all data points corresponding to other targets in the family
were used for training only and the $n_t$ points corresponding to $t$
were $k$-folded with $k=\min(n_t,10)$. That is, for each fold, an SVM
classifier was trained on all points involving other targets of the
family plus a fraction of the points involving $t$, then the
performances of the classifier were tested on the remaining fraction
of data points for $t$. This protocol is intended to assess the
incidence of using ligands from other targets on the accuracy of the
learned classifier for a given target. Second, for each target $t$ we
learned an SVM classifier using only interactions that did not involve
$t$ and tested on the points that involved $t$. This is intended to
simulate the behavior of our framework when making predictions for
orphan targets, \emph{i.e.}, for targets for which no ligand is known.

For the first protocol, since learning an SVM with only one training
point does not really make sense and can lead to "anti-learning" less
than $0.5$ performances, we set all results $r$ involving the Dirac
target kernel on targets with only $1$ known ligand to $\max(r,0.5)$.
This is to avoid any artefactual penalization of the Dirac approach
and make sure we measure the actual improvement brought by sharing
information across targets.

\section{Results}

We first expose the results obtained on the three datasets for the
first experiment, assessing how using training points from other
targets of the family improves prediction accuracy with respect to
individual (Dirac-based) learning.
Table~\ref{tab:exp1} shows the mean success rate across the family
targets for an SVM with a product kernel using the Tanimoto kernel for
ligands and various kernels for proteins.
\begin{figure}[ht]
  \centering
  \begin{minipage}[b]{0.3\linewidth}
  \centering
  \includegraphics[width=\linewidth]{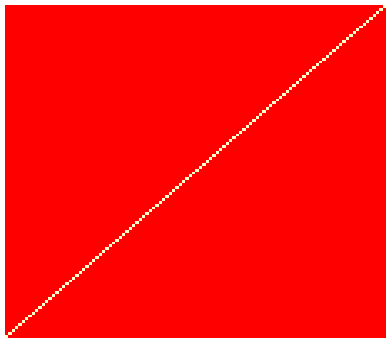}
  \end{minipage}
  \hspace{0.2cm}
  \begin{minipage}[b]{0.3\linewidth}
  \centering
  \includegraphics[width=\linewidth]{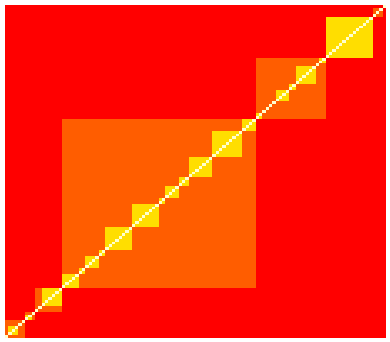}
  \end{minipage}
  \hspace{0.2cm}
  \begin{minipage}[b]{0.3\linewidth}
  \centering
  \includegraphics[width=\linewidth]{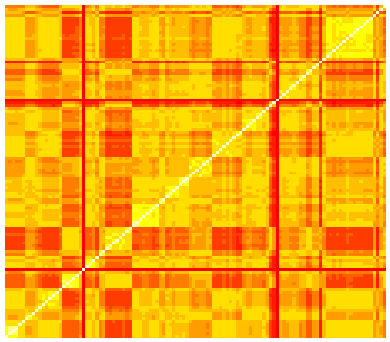}
  \end{minipage}
  \caption{Target kernel Gram matrices ($K_{tar}$) for ion channels
    with multitask, hierarchy and local alignment kernels.}
  \label{fig:kernels}
\end{figure}
For the enzymes and ion channels datasets, we observe significant
improvements when the multitask kernel is used in place of the Dirac
kernel, on the one hand, and when the hierarchy kernel replaces the
multitask kernel, on the other hand. For example, the Dirac kernel
only performs at an average accuracy of $70\%$ for the ion channel
dataset, while the multitask kernel increases the accuracy to $80\%$
and the hierarchy kernel brings it to $88\%$. For the enzymes, a
global improvement of $34.1\%$ is observed between the Dirac and the
hierarchy approaches. This clearly demonstrates the benefits of
sharing information among known ligands of different targets, on the
one hand, and the relevance of incorporating prior information into
the kernels, on the other hand.

On the GPCR dataset though, the multitask kernel performs worse than
the Dirac kernel, probably because some targets in different
subclasses show very different binding behavior which results in
adding more noise than information when sharing naively with this
kernel. However a more careful handling of the similarities between
GPCRs through the hierarchy kernel again results in significant
improvement over the Dirac kernel (from $68.2\%$ to $81.7\%$), again
demonstrating the relevance of the approach.

Sequence-based target kernels do not achieve the same performance as
the hierarchy kernel, although they perform relatively well for the
ion channel dataset. In the case of enzymes, it can be explained by
the diversity of the proteins in the family and for the GPCR, by the
well known fact that the receptors do not share overall sequence
homology~\citep{Gether2000Uncovering}. Figure~\ref{fig:kernels} shows
3 of the tested target kernels for the ion channel dataset. The
hierarchy kernel adds some structure information with respect to the
multitask kernel, which explains the success rate increase. The local
alignment sequence-based kernels fail to precisely re-build this
structure but retains some substructures. In the cases of GPCR and
enzymes, almost no structure is found by the sequence kernels, which,
as alluded to above, was expectable and suggests that more subtle
comparison of the sequences would be required to exploit the
information they contain.

\begin{figure}[ht]
  \begin{minipage}[b]{0.33\linewidth}
  \centering
  \includegraphics[width=\linewidth]{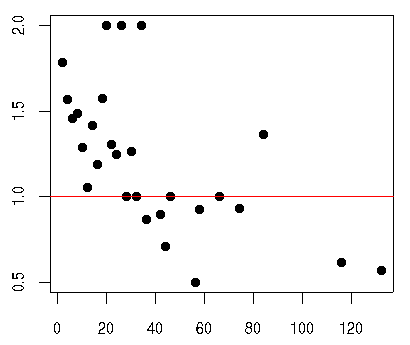}
  \end{minipage}
  \begin{minipage}[b]{0.33\linewidth}
  \centering
  \includegraphics[width=\linewidth]{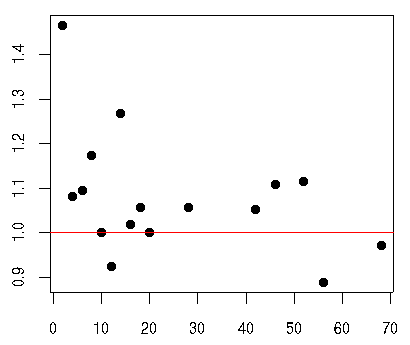}
  \end{minipage}
  \begin{minipage}[b]{0.33\linewidth}
  \centering
  \includegraphics[width=\linewidth]{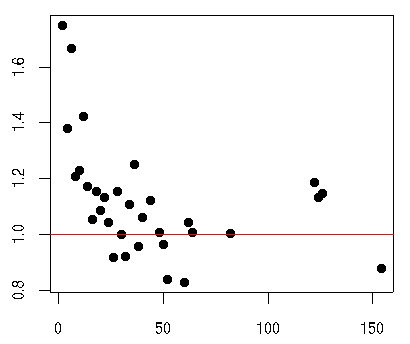}
  \end{minipage}
  \caption{Relative improvement of the \emph{hierarchy} kernel against
    the \emph{Dirac} kernel as a function of the number of known
    ligands for enzymes, GPCR and ion channel datasets.}
  \label{fig:ratios}
\end{figure}

Figure~\ref{fig:ratios} illustrates the influence of the number of
training points for a target on the improvement brought by using
information from similar targets. As one could expect, the improvement
is very strong when few ligands are known and decreases when enough
training points become available. After a certain point (around $30$
training points), using similar targets can even deteriorates the
performances. This suggests that the method could be globally improved
by learning for each target independently how much information should
be shared, for example through kernel learning
approaches~\citep{Lanckriet2004statistical}.

\begin{table}[ht]
  \centering
  \begin{tabular}{*{4}{c}}\hline
    $K_{tar} \backslash$ Target & Enzymes & GPCR & Channels\\\hline
    Dirac  & $0.536\pm0.005$ & $0.682\pm0.022$ &  $0.701\pm0.017$  \\
    multitask & $0.874\pm0.008$ & $0.595\pm0.030$  & $0.797\pm0.017$ \\
    hierarchy & $0.877\pm0.008$ & $0.817\pm0.025$ & $0.857\pm0.015$ \\
    mismatch & $0.582\pm0.008$ & $0.638\pm0.030$ & $0.811\pm0.016$ \\
    local alignment & $0.544\pm0.007$ & $0.696\pm0.033$ & $0.824\pm0.015$ \\\hline
  \end{tabular}
  \caption{Prediction accuracy for the first protocol on each dataset
    with various target kernels.}
  \label{tab:exp1}
\end{table}

The second experiment aims at pushing this remark to its limit by
assessing how each strategy is able to predict ligands for proteins
with no known ligand. Table~\ref{tab:exp2} shows the results in that
case. As expected, the classifiers using Dirac kernels show random
behavior in this case since using a Dirac kernel with no data for the
target amounts to learning with no training data at all. On the other
hand we note that it is still possible to obtain reasonable results
using adequate target kernels. In particular, the hierarchy kernel
loses only $5.2\%$ for the ion channel dataset, $4.1\%$ for the GPCR
dataset and $1.5\%$ compared to the first experiment where known
ligands were used, suggesting that if a target with no known compound
is placed in the hierarchy through, \emph{e.g.} in the case of GPCR
homology detection with known members of the family using specific
GPCR alignment algorithms~\citep{Kratochwil2005Automated} or
fingerprint analysis~\citep{Attwood2003PRINTS}, it is possible to
predict some of its ligands almost as accurately as if some of them
were already available.

\begin{table}[ht]
  \centering
  \begin{tabular}{*{4}{c}}\hline
    $K_{tar} \backslash$ Target & Enzymes & GPCR & Channels\\\hline
    Dirac  &  $0.500\pm0.000$ & $0.500\pm0.000$ & $0.500\pm0.000$ \\
    multitask & $0.856\pm0.009$ & $0.477\pm0.025$ & $0.636\pm0.021$ \\
    hierarchy & $0.862\pm0.009$ & $0.776\pm0.026$ & $0.805\pm0.018$ \\
    mismatch & $0.569\pm0.007$  & $0.579\pm0.028$ & $0.671\pm0.020$ \\
    local alignment & $0.521\pm0.004$ & $0.647\pm0.030$ & $0.722\pm0.019$\\\hline
  \end{tabular}
  \caption{Prediction accuracy for the second protocol on each dataset
    with various target kernels.}
  \label{tab:exp2}
\end{table}

\section{Discussion}

We propose a general method to combine the chemical and the biological
space in a principled way and predict interaction between any small
molecule and any target, which makes it a vary valuable tool for drug
discovery. The method allows to represent systematically a
ligand-target couple, including information on the interaction between
the ligand and the target. Prediction is then performed by any machine
learning algorithm (an SVM in our case) in the joint space, which
makes targets with few known ligands benefit from the data points of
similar targets, and which allows to make predictions for targets with
no known ligand. Our information sharing process therefore simply
relies on a description choice for the ligands, another one for the
targets and on classical machine learning methods: everything is done
by casting the problem in a joint space and no explicit procedure to
select which part of the information is shared is needed. Since it
subdivides the representation problem into two subproblems, our
approach makes use of previous work on kernels for molecular graphs
and kernels for biological targets. For the same reason, it will
automatically benefit from future improvements in both fields.  This
leaves plenty of room to increase the performance.

Results on experimental ligand datasets show that using target kernels
allowing to share information across the targets considerably improve
the prediction, especially in the case of targets with few known
ligands. The improvement is particularly strong when the target kernel
uses prior information on the structure between the targets,
\emph{e.g.}, a hierarchy defined on a target class. Although sequence
kernels did not give very good results in our experiments, we believe
using the target sequence information could be an interesting
alternative or complement to the hierarchy kernel. Further improvement
could come from the use of kernel for structures in the cases where 3D
structure information is available (\emph{e.g.} for the enzymes, but
not for the GPCR). Our method also shows good performances even when
no ligand at all is known for a given target, which is excellent news
since classical ligand based approaches fail to predict ligand for
these targets in the one hand, and docking approaches are
computationally expensive and not feasible when the target 3D
structure is unknown which is the case of GPCR in the other hand.

In future work, it could be interesting to apply this framework to
quantitative prediction of binding affinity using regression methods
in the joint space. It would also be important to confirm predicted
ligands experimentally or at least by docking approaches when the
target 3D structure is available.

\section*{Acknowledgments}
We thank Pierre Mah\'e for his help with ChemCPP and kernels for
molecules, and V\'eronique Stoven for insightful discussions on the
biological aspects of the problem.

\end{document}